\documentclass[hidelinks,twocolumn, prl, superscriptaddress,a4paper]{revtex4}

\usepackage[colorlinks=true,allcolors=blue]{hyperref}
\usepackage{graphicx}
\usepackage{dcolumn}
\usepackage{orcidlink}
\usepackage{csquotes}
\usepackage{float}
\usepackage{amsmath,amssymb}
\usepackage{bm}
\usepackage{braket}
\usepackage{xcolor}
\usepackage{comment}

\begin{document}

\preprint{APS123-QED}



\author{Maitri Ganguli\,\orcidlink{0009-0009-4701-2459}}
\affiliation{Department of Physics, Indian Institute of Science, Bangalore}


\author{Aneek Jana\,\orcidlink{0009-0001-1097-4250}}

\affiliation{Undergraduate Program, Indian Institute of Science, Bangalore}

\author{Awadhesh Narayan\,\orcidlink{0000-0003-0127-7047}}

\affiliation{Solid State and Structural Chemistry Unit, Indian Institute of Science, Bangalore}

\title{Tunable topology, Hall response, and spin-textures in bicircularly polarized light illuminated altermagnets}



\date{\today}

\begin{abstract}
Altermagnets, featuring non-relativistic spin splitting, have drawn enormous attention due to their intriguing properties. Here, we investigate the effects of shining bicircularly polarized light (BCL) on altermagnets with Rashba spin-orbit coupling. We discover a remarkable tunability of topology, spin-textures, and Fermi surfaces of altermagnets by means of BCL illumination, going beyond monochromatic light. We illustrate a cascade of topological phase transitions controllable by BCL and demonstrate how these transitions are reflected in the anomalous Hall response of the altermagnet. Furthermore, we show that the spin-textures and Fermi surfaces can be directly tuned by the relative phase of the BCL, stemming from the underlying symmetry changes. Our findings can pave the way for effectively controlling altermagnetic materials with structured light.
\end{abstract}


\maketitle

\textit{Introduction.} 
For the past decade, Floquet engineering using monochromatic circularly polarized light (CL) has emerged as a promising approach to control quantum matter~\cite{oka2009photovoltaic,lindner2011floquet,Wang_2014,CL1,CL2,CL3,CL4,CL5}. Recently, bicircularly polarized light (BCL) has come to fore as a notable advance to tune various properties of materials. Due to the superposition of two light beams with multiple tuning parameters, BCL allows a degree of control that is not possible with monochromatic light. The interplay of two light beams can yield enriched physics, including enhanced tunability of band structure, topology, symmetry, and photoresponse, to name just a few~\cite{BCLPRL,bcl1,bcl2,bcl4,bcl5,mrudul2021controlling,strobel2023linear,Ganguli_2025,kanega2025two,guo2025bicircular}.

In recent years, there have been several studies indicating a new classification of magnets with a mixed behavior of ferromagnets and antiferromagnets, namely, they have spin-split bands despite having vanishing net magnetization -- these have been termed altermagnets~\cite{altermagnet1,altermagnet2,altermagnet3}. In such systems, time-reversal symmetry by itself is broken, but time-reversal combined with a lattice point-group symmetry (other than translation or inversion) is still preserved, resulting in spin-split bands with vanishing net magnetization. Owing to the spin-split bands, altermagnets display intriguing Fermi surfaces and spin-textures, which have also been recently starting to be measured~\cite{Yang_2025,PhysRevB.111.125119,Bai_2024,5js8-2hj8,lai2025dwaveflatfermisurface}.

\begin{figure}[b]
    \centering
    \includegraphics[width=0.49\linewidth]{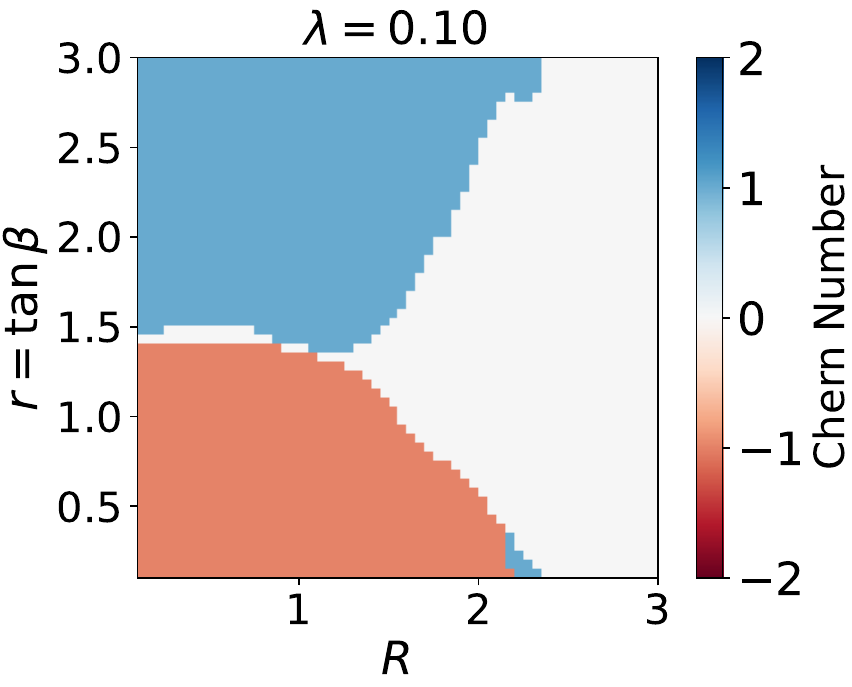} 
    \includegraphics[width=0.49\linewidth]{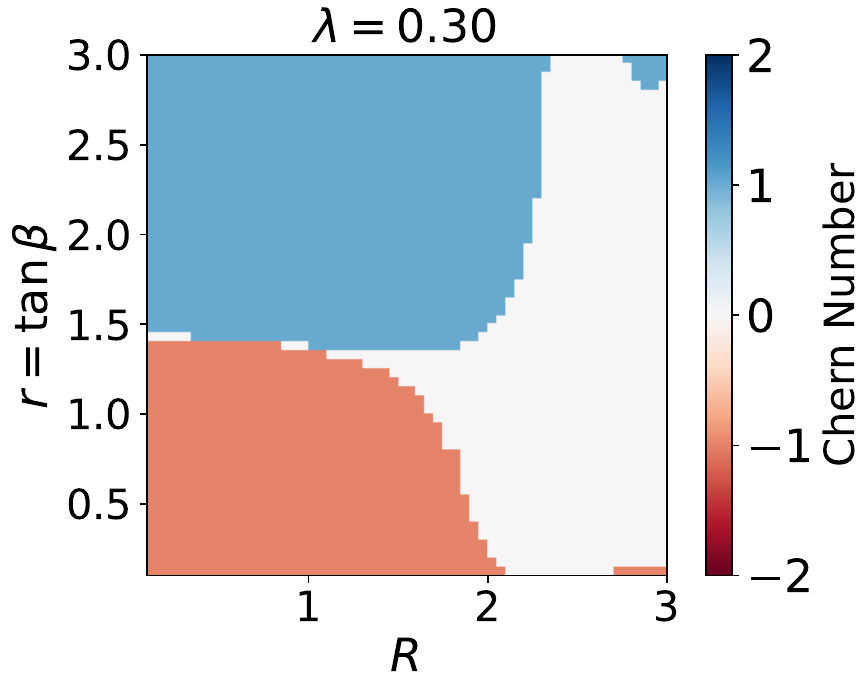}

    \includegraphics[width=0.49\linewidth]{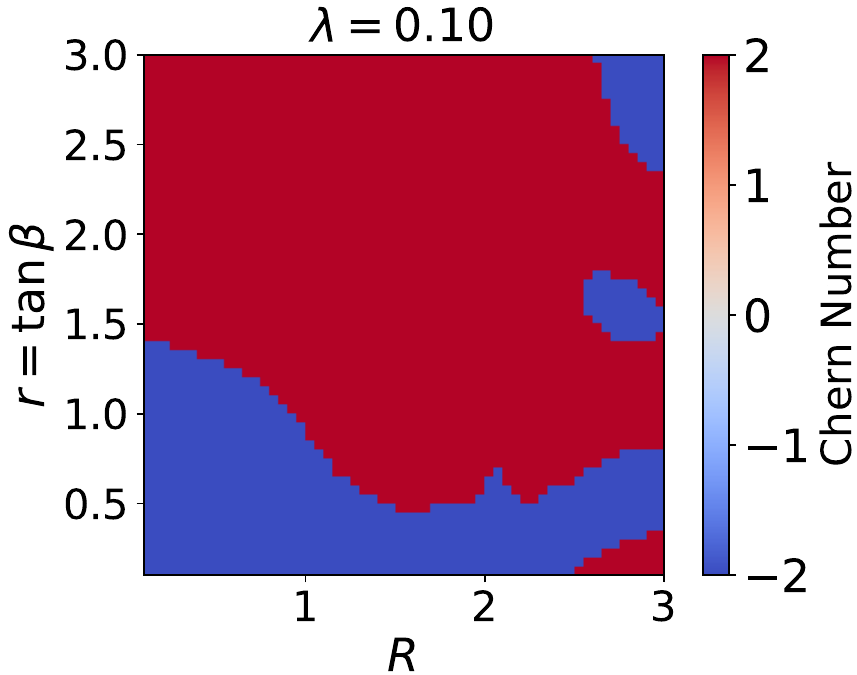} 
    \includegraphics[width=0.49\linewidth]{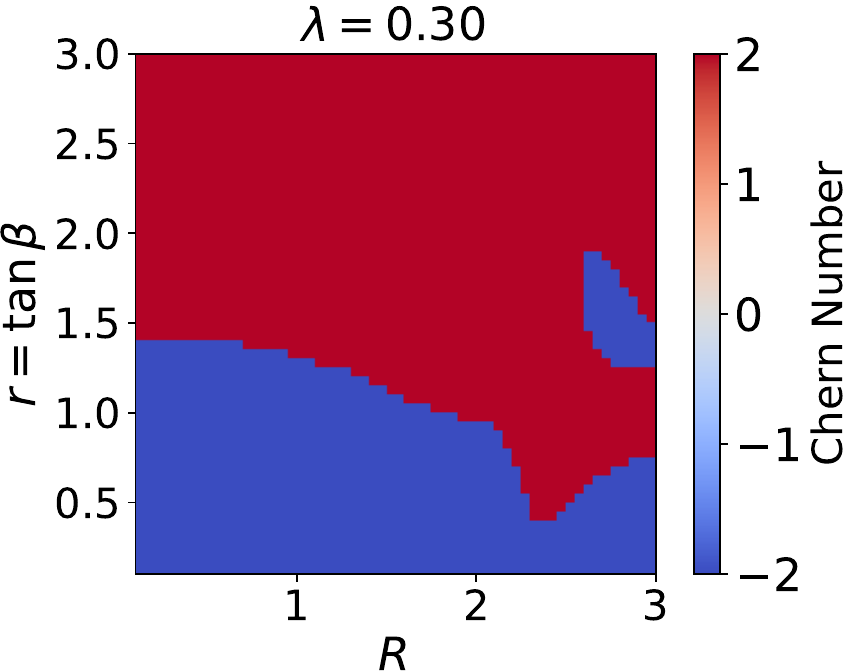}

    \caption{\textbf{Chern number phase diagram for BCL illuminated \(\bm d\)-wave altermagnets.} The plots illustrate topological phases obtained with varying relative amplitude \(r\) and overall amplitude \(R\). The top panel is for \(d_{x^2-y^2}\)-wave altermagnets and the bottom panel is for \(d_{xy}\)-wave altermagnets. The strengths of RSOC, $\lambda$, are indicated in the plots. Other parameters used are \(t_j=0.3, \omega=3,\) and \(\alpha=0\). This showcases a direct control over the Chern number and topological phases by BCL illumination.}
    \label{fig: chern number density plots}
\end{figure}

In this work, we probe the effects of BCL illumination on altermagnets to reveal a tunable topology as well as controllable spin-textures, and Fermi surfaces. We highlight a variety of topological phase transitions directly triggered by BCL illumination and examine the resulting effects in anomalous Hall response of the system. We reveal that spin-textures and Fermi surfaces can be tuned by the relative phase of the BCL, which allows control over the symmetry of the altermagnet. Overall, we showcase that BCL allows a remarkable control over altermagnets, going beyond monochromatic light.

\textit{Altermagnet models and BCL tuning.}
We take the following second-quantized Hamiltonian for a \(d_{x^2-y^2}\)-wave altermagnet on a square lattice (with primitive lattice vectors \(\hat{\bm x}\) and \(\hat{\bm y}\), with lattice spacing set to one for simplicity),

\begin{equation}
   \begin{aligned}
   &\hat{\mathcal H}_{\text{AM}} = \sum_{\bm k}c_{\bm k}^\dagger H_{\text{AM}} (\bm{k}) c_{\bm k}-\mu c_{\bm k}^\dagger c_{\bm k}, \,\,\quad c_{\bm k}^\dagger = (c_{\bm k\uparrow}^\dagger\quad c_{\bm k\downarrow}^\dagger),\\   
    &H_{\text{AM}}^{d_{x^2-y^2}} (\bm{k}) = - 2 t (\cos k_x + \cos k_y) I + 2 t_j (\cos k_y-\cos k_x)\sigma_z,
    \end{aligned}
\end{equation} 

where \(c_{\bm k,\uparrow(\downarrow)}^\dagger\) is the creation operator of up (down) spin at crystal momentum \(\bm k\), \(\mu\) is the chemical potential,and \(t+ t_j\) and \(t- t_j\) are the (real) hopping amplitudes of up-spin (down-spin) along \(\hat{\bm{x}}\) (\(\hat{\bm{y}}\)) and \(\hat{\bm{y}}\) (\(\hat{\bm{x}}\)) directions, respectively. Due to the spin-diagonal nature of the Hamiltonian, the Berry curvature \(\Omega (\bm k)=i(\bra{\partial_{k_x}u(\bm k)}\partial_{k_y}u(\bm k)\rangle-\bra{\partial_{k_y}u(\bm k)}\partial_{k_x}u(\bm k)\rangle)\), vanishes identically as the eigenstates \(\ket{u(\bm k)}\) are independent of \(\bm k\). This remains true even if light is illuminated on the system. Therefore, this simplistic model cannot give rise to the anomalous Hall effect (AHE)~\cite{AHE,QAHE}, even in the presence of illumination. Next, we introduce a nearest neighbor Rashba-type spin-orbit coupling (RSOC) with coupling constant $\lambda$~\cite{Rashba},

\begin{equation}
    H_{\text{RSOC}}(\bm k) = 2 \lambda (\sin k_y \sigma_x - \sin k_x \sigma_y).
\end{equation}

Now, in the full Hamiltonian \(H(\bm k) = H_{\text{AM}}^{d_{x^2-y^2}}(\bm k)+H_{\text{RSOC}}(\bm k)\), the degeneracies along \(\Gamma-M\) line are lifted, but at the points \(\Gamma, M\) they remain intact. The addition of the Rashba term generates a non-vanishing Berry curvature, leading to the possibility of AHE. However, with our choice of out-of-plane magnetic order, the combined rotation-time-reversal invariance forbids AHE~\cite{KnolleTubaleTopology}. In this case, however, it is possible to open up a band-gap by illumination with light, which can give rise to topologically non-trivial bands. As we will see, we obtain a finite AHE by the application of light. 

In the above-described model, the spin-asymmetric hopping direction (\(\hat{\bm x}\) and \(\hat{\bm y}\)) coincide with the direction of RSOC. Next, we also consider a model for \(d_{xy}\)-wave altermagnet that has spin-asymmetric hopping along diagonal neighbors (\(\hat{\bm x}+\hat{\bm y}\) and \(\hat{\bm x}-\hat{\bm y}\)), keeping the RSOC in the same direction. Here, the altermagnetic part of the Hamiltonian can be written as,

\begin{equation}
    H_{\text{AM}}^{d_{xy}}(\bm k) = - 2 t(\cos k_x + \cos k_y) I + 2 t_j \sin k_x \sin k_y \sigma_z.
\end{equation}

In this case, in the presence of RSOC we have four gapless points in the Brillouin zone at \(\Gamma, M, X\)  and \(Y\). We will see in our analyses how the responses in the presence of laser irradiation differ for these two different relative configurations of spin-orbit coupling and altermagnetic exchange coupling.

\begin{figure}
    \centering
    \includegraphics[width=0.48\linewidth]{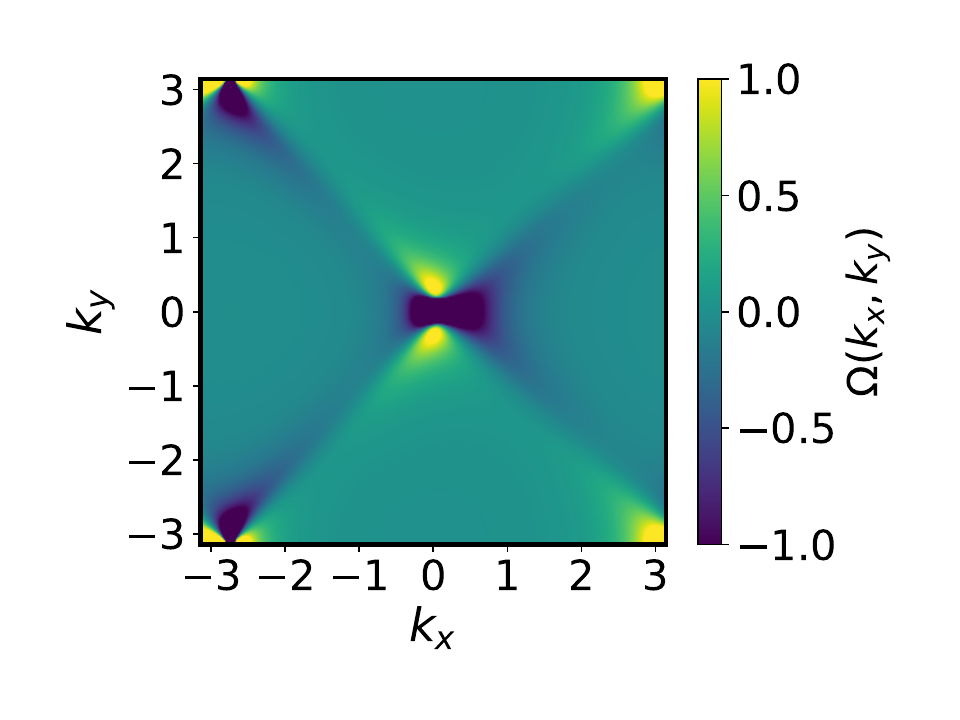}
    \includegraphics[width=0.48\linewidth]{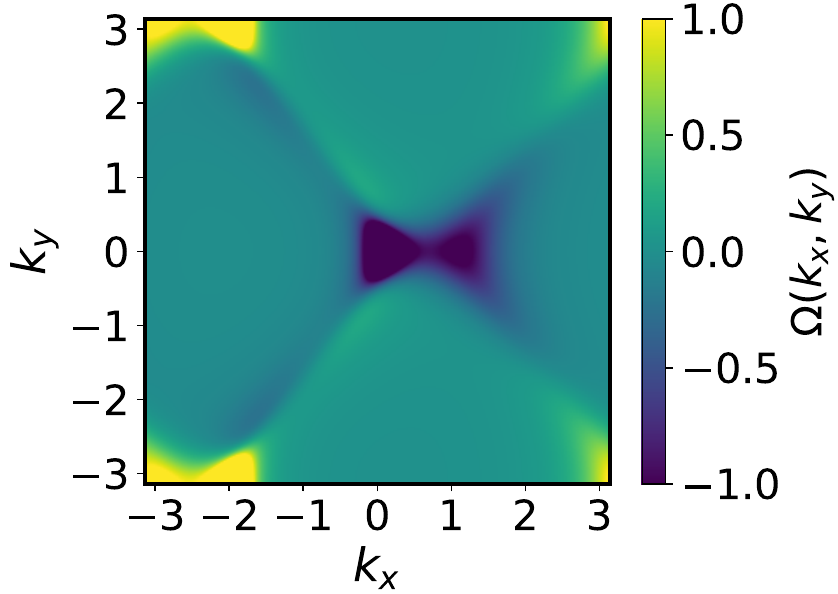}
    \includegraphics[width=0.48\linewidth]{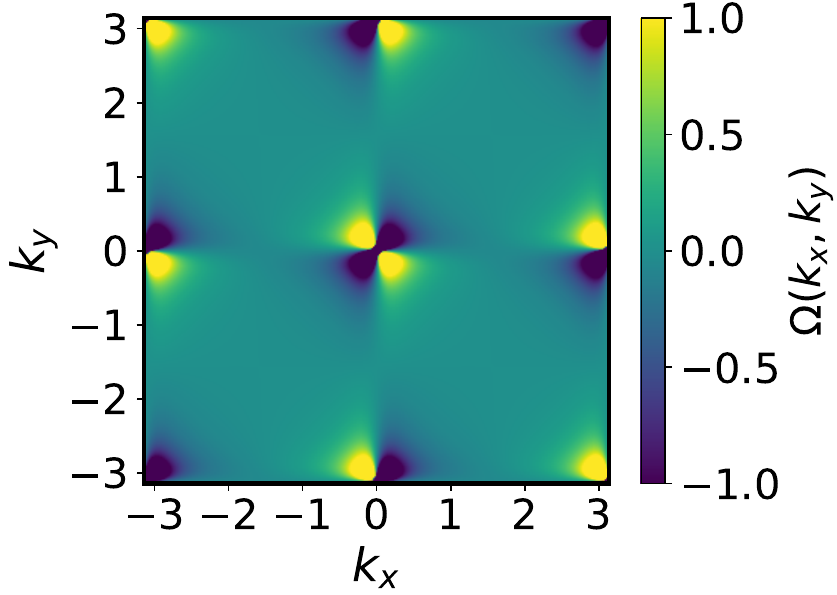}
    \includegraphics[width=0.48\linewidth]{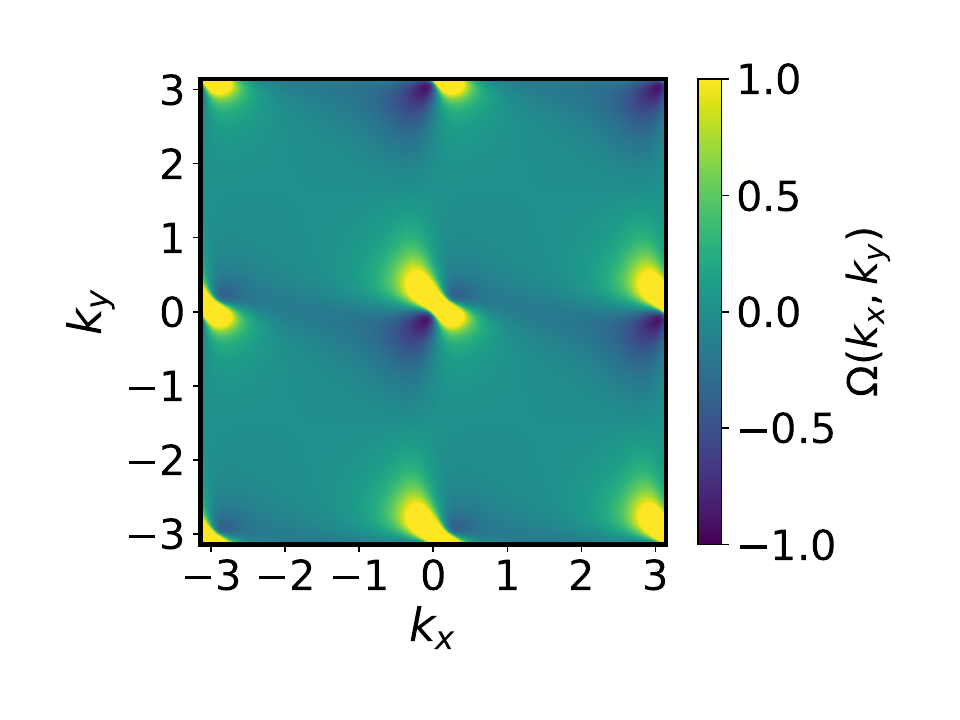}
    
    \caption{\textbf{Overall amplitude driven topological phase transition in \(\bm d\)-wave altermagnets.} Berry curvature distribution in the Brillouin zone for different values of overall amplitude $R$. For \(d_{x^2-y^2}\)-wave altermagnets (top panel), increasing the overall amplitude \(R\) at a fixed relative amplitude \(r\) can lead to a transition from topologically non-trivial phase with \(C=-1\) at \(R=1.5\) (top-left) to a topologically trivial phase with \(C=0\) at \(R=2.0\) (top-right). For \(d_{xy}\)-wave altermagnets (bottom panel), however, the transition takes place between two topologically non-trivial phases, for instance, from \(C=-2\) at \(R=0.5\) (bottom-left) to \(C=+2\) at \(R=1.2\) (bottom-right). Here we have fixed \(r=1,  \omega=3,\alpha=0\) for \(t_j=0.3,\lambda=0.1\).}
    \label{fig: berry curvature at diff overall amplitude}
\end{figure}

We next incorporate the effect of illumination by BCL and write the vector potential as~\cite{Ganguli_2025},

\begin{equation}
\begin{aligned}
    &A_x = A_0 (r\cos(\eta\omega t -\alpha)+\cos(\omega t)),\\
    & A_y = A_0 (-r\sin(\eta\omega t -\alpha)+\sin(\omega t)).
\end{aligned}
\end{equation}

Here $A_0$ is the amplitude controlled by the intensity of the light, $r$ is the relative amplitude of the two beams, $\omega$ is frequency of one of the components, and $\eta$ and $\alpha$ are the relative frequency and relative phase difference between the two components of the BCL beam. Note that \(r\rightarrow 0\) recovers CL from BCL. In this work, we will mainly work with \(\eta=2\), i.e, the frequency of one beam is twice that of the other. We further choose a different parametrization such that \(A_0 r \rightarrow R \sin\beta, \,\, A_0\rightarrow R\cos\beta\), where \(R\) has the interpretation of an overall amplitude and \(r=\tan\beta\) is a relative amplitude.

We use Floquet theory to derive an effective Hamiltonian which describes the dynamics of the system at stroboscopic time-scales~\cite{Floquet2,Floquet1,BW_HFE,Floquet3}. We determine the time-periodic Hamiltonian by means of the Peierls substitution \(k_x \rightarrow k_x + A_x (t)\) and \(k_y \rightarrow k_y + A_y (t)\) (in the units \(e=1\)) in the Bloch Hamiltonian, yielding \(H(\bm k,t) = H(\bm k+ \bm A(t))\). The Floquet effective Hamiltonian is given in terms of the Fourier modes of the time-periodic Hamiltonian,

\begin{equation}
    H_F (\bm k) \approx H_0  +\sum_{n>0}\left(\frac{[H_{+n},H_{-n}]}{n\omega}+ \frac{[H_{0},H_{+n}]-[H_{0},H_{-n}]}{2 n\omega}\right),
\end{equation}

where, \(H(\bm k,t) = \sum_m H_m (\bm k) e^{i m \omega t}\).

Next, we consider the illumination of altermagnetic models with the BCL. Writing the Fourier components as \(H_m(\bm k) = d_m^0(\bm k) I + \bm d_m(\bm k)\cdot \bm \sigma\), the effective Hamiltonian takes the form~\cite{SM1},

\begin{equation}
\begin{aligned}
    & H_F(\bm{k})\\
    &\approx d^0_0(\bm{k}) I + \bm d_0(\bm{k})\cdot \bm \sigma\\
    &-\frac{2}{\omega}\sum_{m\in\mathbb Z+} \frac{\text{Im}(\bm d_m (\bm{k})\times \bm d_m (\bm{k})^*+\bm d_0 (\bm{k})\times \bm d_m (\bm{k}))}{m}\cdot \bm \sigma\\
    &= d^0_F (\bm{k}) I + \bm{d}_F(\bm{k})\cdot \bm\sigma.
\end{aligned}
\end{equation}

With the effective Floquet Hamiltonian, \(H_F(\bm k)\), we next analyze the band structures, topological phase transitions, spin textures, Fermi surfaces, and Berry curvature of the altermagnets with different BCL and model parameters. In passing, we note that the band topology, Berry curvature and spin-texture are completely determined by the vector \(\bm d_F(\bm k)\), while the Fermi surface can be crucially affected by \(d_F^0(\bm k)\) too. The spin-texture is the expectation value of \(\bm \sigma\) in the eigenstates of \(H_F (\bm k)\), which for the lower band is given by $-\hat{\bm d}_F(\bm k) = - \frac{\bm d_F(\bm k)}{|\bm d_F(\bm k)|}$. On the other hand, the Berry curvature of the bands can be determined from

\begin{equation}
    \Omega_\pm(\bm k) = \mp \frac{1}{2 |\bm d_F(\bm k)|^3} \bm d_F(\bm k)\cdot (\partial_{k_x}\bm d_F(\bm k) \times \partial_{k_y}\bm d_F(\bm k)),
\end{equation}

where \(\pm\) indicate upper (lower) band. When the bands are non-degenerate at each \(\bm k\) (i.e., \(|\bm d_F(\bm k)|\neq 0\) throughout the Brillouin zone), one can define the Chern number utilizing the Berry curvature, \(C=(1/2\pi)\int_{BZ} d^2\bm k\,\, \Omega(\bm k) \) (for the lower band). Experimentally observable signature of non-trivial Berry curvature is the anomalous Hall conductivity \(\sigma_{\text{AH}}\)~\cite{AHE}, which is the average of the Berry curvature weighted by the Fermi function, dependent on filling or chemical potential (\(\mu\)) and temperature (\(T\)), given by 

\begin{equation}
    \begin{aligned}
       \sigma_{\text{AH}}= -\frac{e^2}{\hbar} \int \frac{d^2\bm k}{(2\pi)^2} (n_{F+}(\bm k)-n_{F-}(\bm k))\Omega_- (\bm k).
    \end{aligned}
\end{equation}

Here \(n_F(x,T)=1/(1+\exp\{x/T\})\) is the equilibrium Fermi distribution function, and \(n_{F\pm}(\bm k) = n_F (E_{\bm k,\pm}-\mu,T)\) where \(E_{\bm k,\pm}\) are the two eigenvalues of the Bloch Hamiltonian at crystal momentum \(\bm k\). Depending on the filling the chemical potential \(\mu\) can be determined. We work in the \(T\rightarrow 0\) limit, where the Fermi function can be approximated by the Heaviside step function, \(\lim_{T\rightarrow 0} n_F(x,T) = \Theta (-x)\).

\begin{figure}[t]
    \centering
    \includegraphics[width=0.48\linewidth]{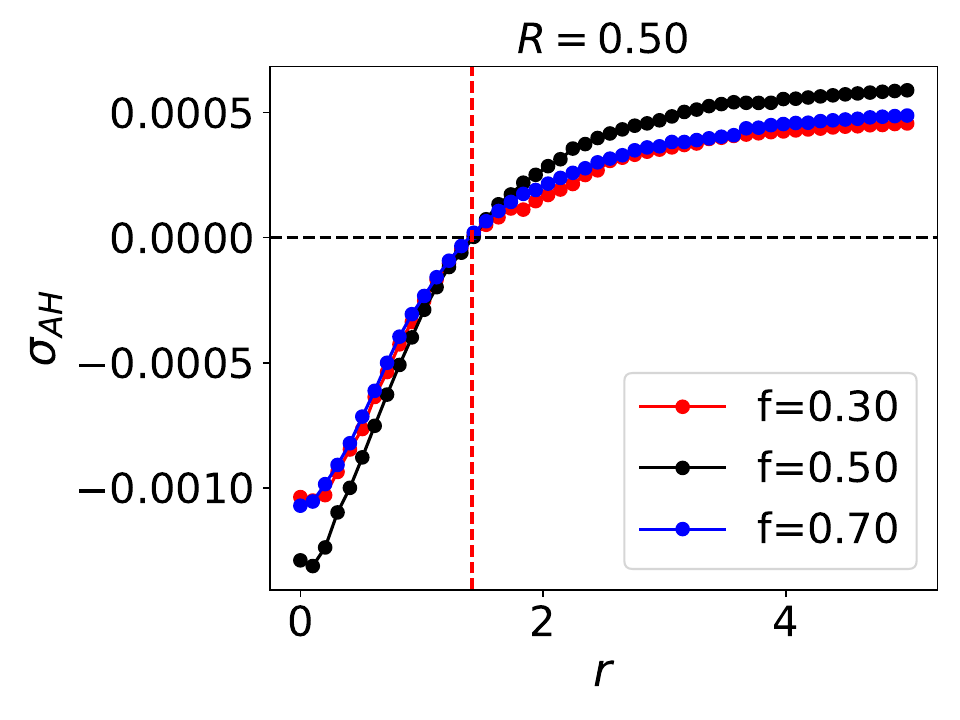}
    \includegraphics[width=0.48\linewidth]{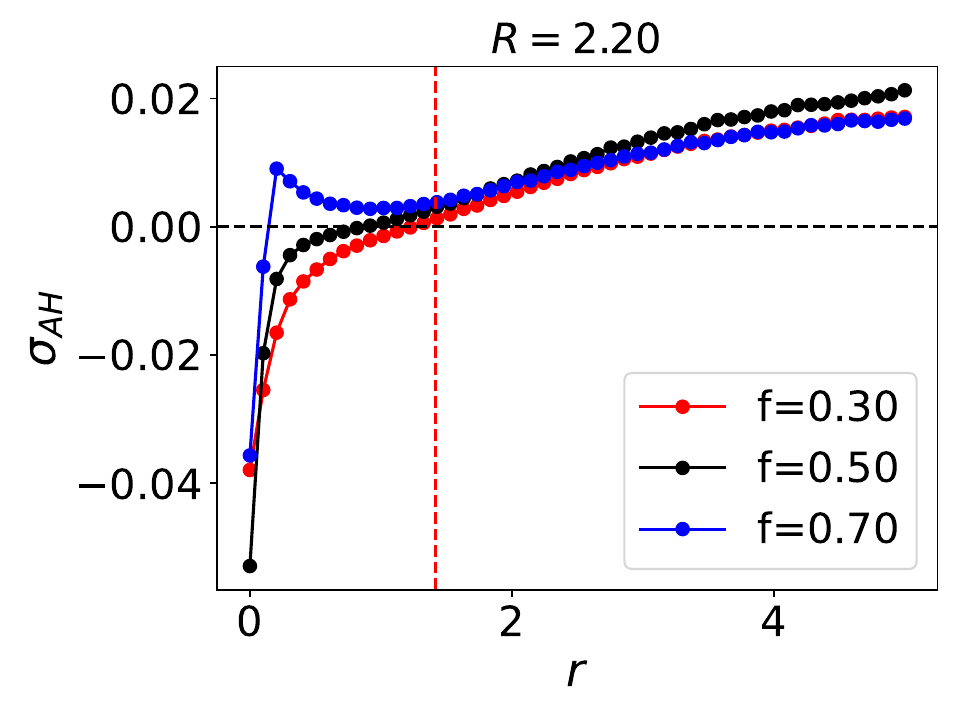}

    \includegraphics[width=0.47\linewidth]{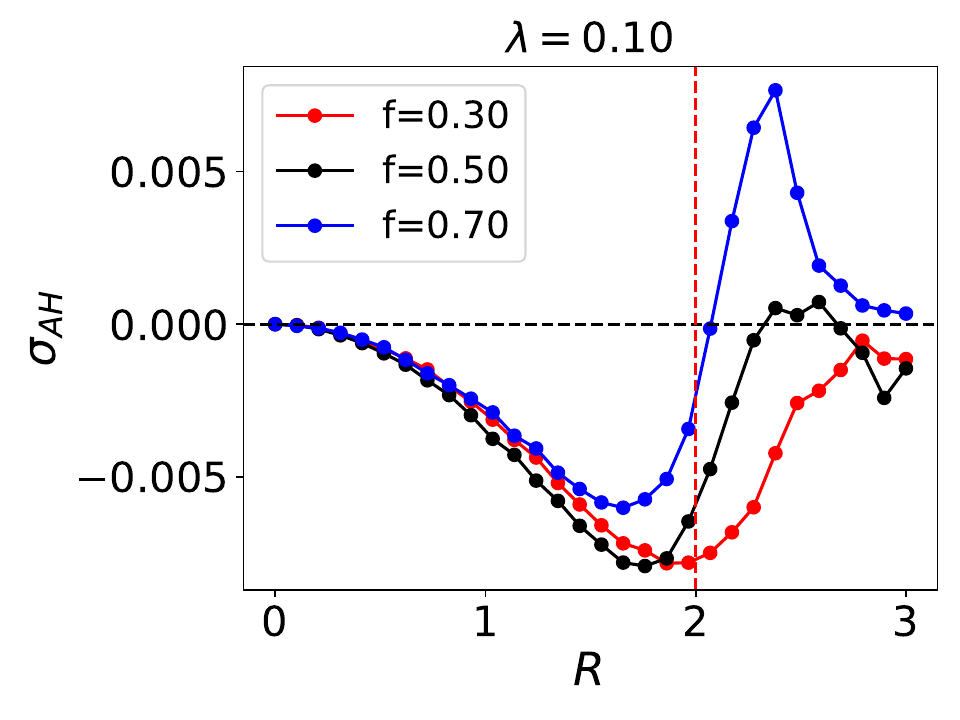}
    \includegraphics[width=0.47\linewidth]{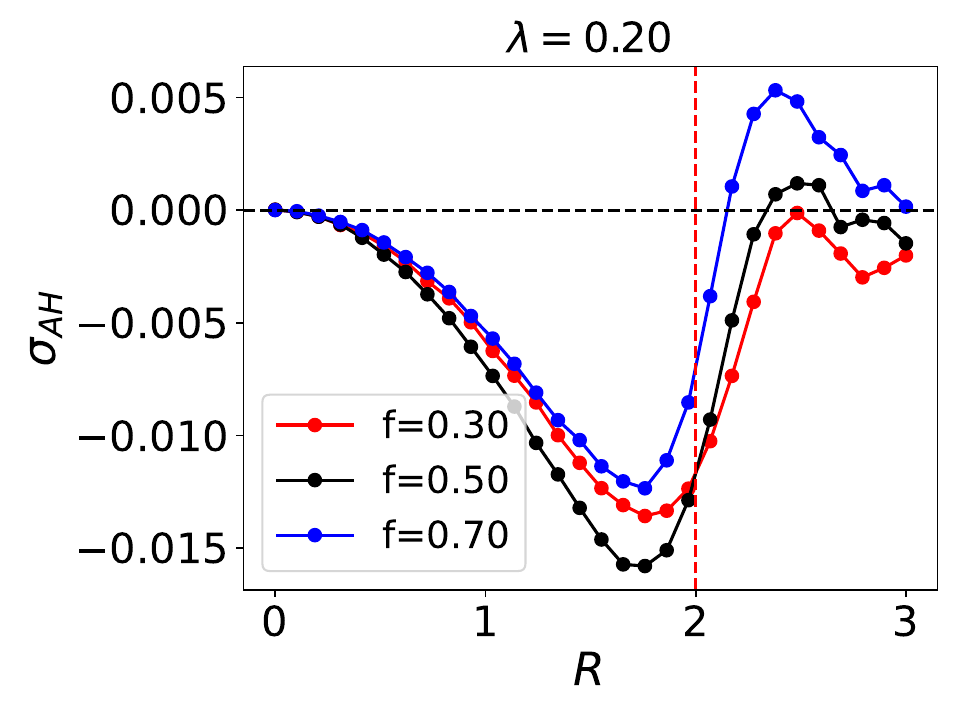}
    \caption{\textbf{Anomalous Hall conductivity as a signature of topological phase transitions.} The top panel shows the topological phase transition by changing the relative amplitude \(r\) as apparent in the sign of the anomalous Hall conductivity \(\sigma_{\text{AH}}\) (in the units of \(e^2/\hbar\)) for different overall amplitude values, \(R\), and at different filling fractions, \(f\), with a fixed \(\lambda=0.1\). This should be understood in conjunction with the Chern number phase diagram in Fig.~\ref{fig: chern number density plots}. The bottom panel shows the topological transition by changing \(R\) at fixed \(r=0.5\). This transition happens through a band inversion around the \(M\) point and the transition is more pronounced for higher filling fractions. The other parameters are set to \(t=1,t_j=0.3,\omega=3, \text{ and } \alpha=0\), and we have chosen a \(d_{x^2-y^2}\)-wave altermagnet.}
    \label{fig: ahe dx2-y2}
\end{figure}

\textit{Tunable Berry curvature and topology.} We begin by presenting the tuning the Berry curvature of the bands by changing parameters of the applied BCL. As we mentioned earlier, a non-zero Berry curvature can be observed in the bands as soon as the Rashba coupling is turned on (\(\lambda\neq 0\)). However, the two bands touch each other at \(\Gamma\) and \(M\) points for \(d_{x^2-y^2}\)-wave altermagnet and at \(\Gamma,X,Y,\) and \(M\) points for \(d_{xy}\)-wave altermagnet, where the Berry curvature is singular. Illumination with light adds effective mass terms at these band-touching points (BTPs) opening up gaps. This smoothens the Berry curvature at these high symmetry points and Chern number becomes well-defined. Now, if we focus on the lower band, we find that the dominant sign of the Berry curvature or the sign of the Chern number crucially depends on the direction of polarization of the CL. For low-enough light amplitude (\(R\ll 1\)), we find that the Chern number is negative for right-circular polarization (\(r\rightarrow 0\)) and positive for left-circular polarization (\(r \rightarrow \infty\)), see Fig.\ref{fig: chern number density plots},  as the sign of the effective mass term crucially depends on the polarization.

In the case of \(d_{x^2-y^2}\)-wave altermagnet and not too high RSOC compared to the altermagnetic exchange coupling (\(\lambda \leq t_j\) or \(\lambda\) slightly greater than \(t_j\)), we find that BCL induces topological phases with possible Chern numbers \(\pm 1\) with \(\pm 0.5\) contribution each from around the \(\Gamma\) and \(M\) points. For the \(d_{xy}\)-wave case, however, we get possible Chern numbers of \(\pm 2\), with \(\pm 0.5\) contributions from around each of four high symmetry points in the two-dimensional Brillouin zone. We have confirmed these results both for the lattice model as well as by analyzing the low-energy effective Hamiltonian around each of those points.

Next we focus on the Chern number phase diagram of illuminated altermagnets to understand how the band topology depends on the overall amplitude \(R\) and the relative amplitude \(r\), as shown in Fig.\ref{fig: chern number density plots}. In the \(R \ll 1\) regime, we find a jump in the Chern number around \(r \approx \sqrt{2}\) (a jump \(|\Delta C|\) =2 for \(d_{x^2-y^2}\)-wave and \(|\Delta C| = 4\) for \(d_{xy}\)-wave). For larger values of \(R\), the transition may occur at an \(r\) different from \(\sqrt{2}\). This topological phase transition occurs via a closing of the band gap, which implies the emergence of BTP even in the presence of non-zero illumination \(R\neq 0\)~\cite{SM1}, similar to what was previously reported for multifold semimetals and Dirac nodal line semimetals~\cite{Ganguli_2025}. This phase transition is also apparent from the Berry curvature plot around the \(\Gamma\) point for different \(r\) at fixed \(R\), for both kinds of \(d\)-wave altermagnets.

We further find that there can also be an overall amplitude-driven topological phase transition at fixed \(r\). This regime of the phase diagram is accessible for single frequency or CL (\(r=0\)) as well, and the case of CL has very recently been investigated in~\cite{altermagnet_cl,altermagnet_cl2,yarmohammadi2025anisotropic}. For the \(d_{x^2-y^2}\)-wave altermagnet, one can drive the system into a topologically trivial phase (with vanishing Chern number, \(C=0\)) by increasing \(R\), unlike the \(d_{xy}\)-wave case, which always remains topologically non-trivial under illumination. The vanishing of the Chern number can be understood from the Berry curvature profile as we increase \(R\), as presented in the top panel of Fig.~\ref{fig: berry curvature at diff overall amplitude}. One can observe a band inversion around the \(M\) point so that its contribution cancels that around the \(\Gamma\) point, rendering the system into a topologically trivial phase. Intriguingly, we find that by increasing \(r\) it is possible to decrease the critical value of overall amplitude \(R_c\) when the topologically non-trivial to trivial phase transition occurs. In contrast, the overall amplitude-driven phase transition in a \(d_{xy}\)-wave altermagnet occurs between two topologically non-trivial phases with opposite Chern numbers, see bottom-panel of Fig.~\ref{fig: berry curvature at diff overall amplitude}. These transitions also happen through a gap closing at one or more points in the Brillouin zone.

Finally, we turn to possible experimental signatures to detect our predicted topological phase transitions. We propose that measuring the anomalous Hall conductivity \(\sigma_{\text{AH}}\) at different filling fractions can directly exhibit the effects of the topological phase transitions that stem from band inversions. The top panel of Fig.~\ref{fig: ahe dx2-y2} shows \(\sigma_{\text{AH}}\) in the BCL irradiated \(d_{x^2-y^2}\)-wave altermagnet as a function of relative amplitude \(r\) for two different overall amplitudes. For small light amplitude it is clear the topological transition around \(r=\sqrt{2}\) is reflected in the sign-change of \(\sigma_{\text{AH}}\). For larger overall amplitude the transitions happen away from \(r=\sqrt{2}\), see Fig.~\ref{fig: chern number density plots}, as also reflected in Fig.~\ref{fig: ahe dx2-y2}. The bottom panel shows \(\sigma_{\text{AH}}\) with overall amplitude \(R\) for a fixed \(r\). The transitions observed here also correlate to the transitions observed in the Chern number density plots. We note that, since the system is metallic,\(\sigma_{\text{AH}}\) is not quantized. Nevertheless, both Chern number and \(\sigma_{\text{AH}}\) showcase the topological phase transitions in a consistent manner.

\begin{figure}[t]
    \centering
    \includegraphics[width=0.31\linewidth]{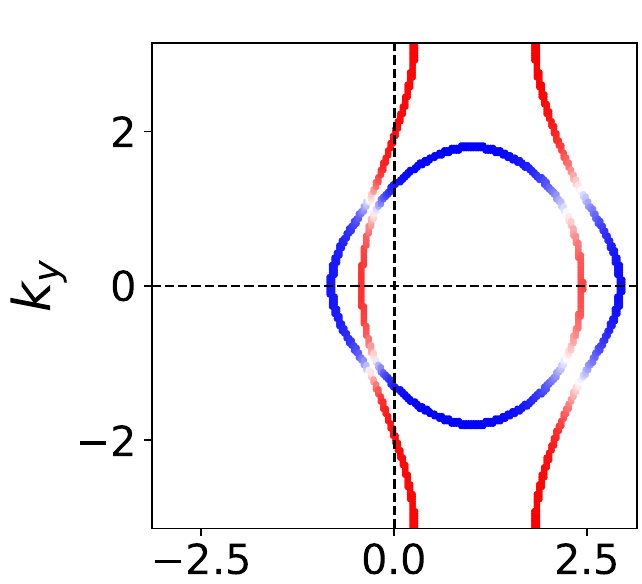}
    \includegraphics[width=0.27\linewidth]{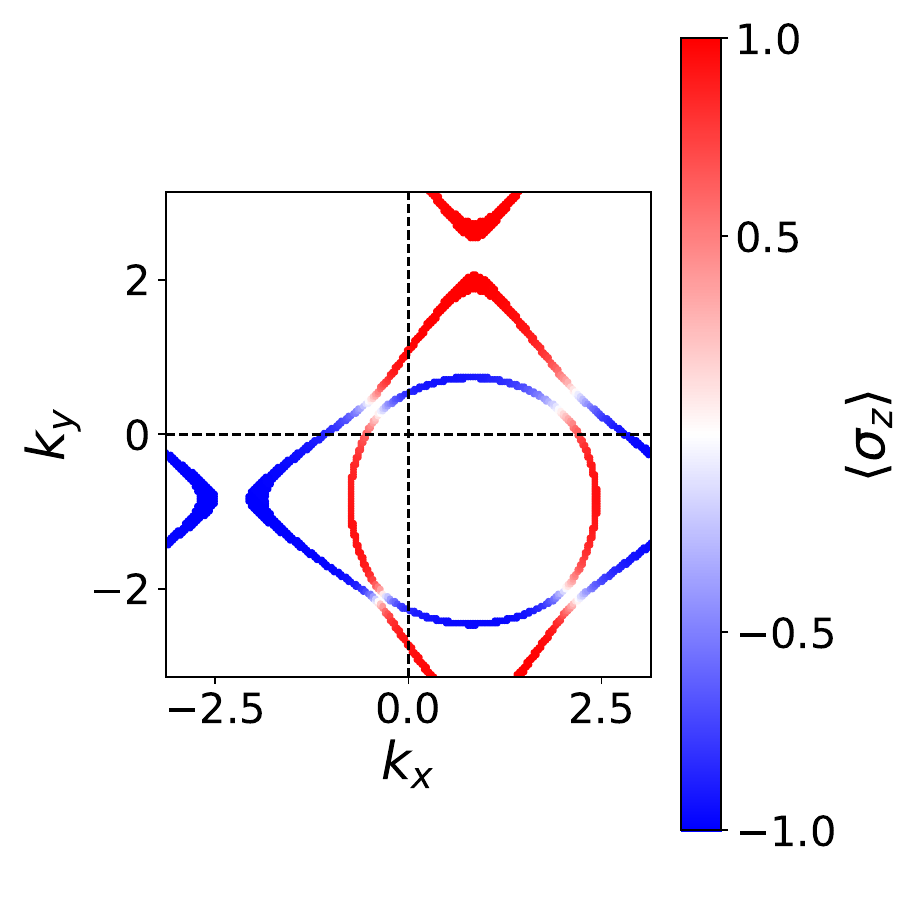}
    \includegraphics[width=0.27\linewidth]{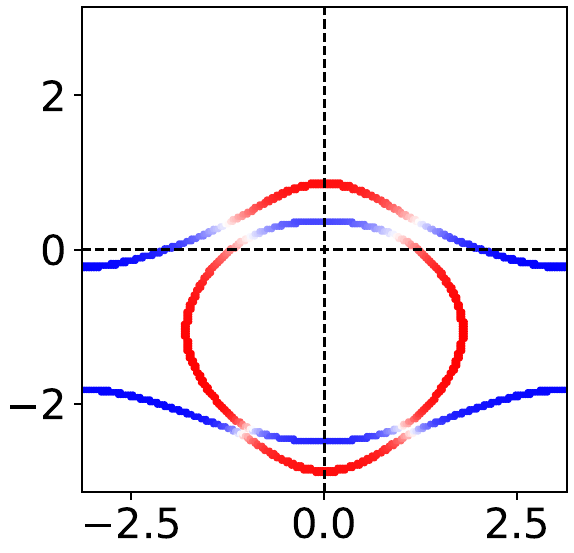}
    \includegraphics[width=0.08\linewidth]{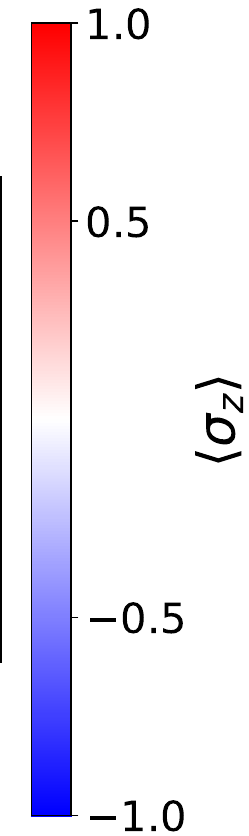}
\includegraphics[width=0.31\linewidth]{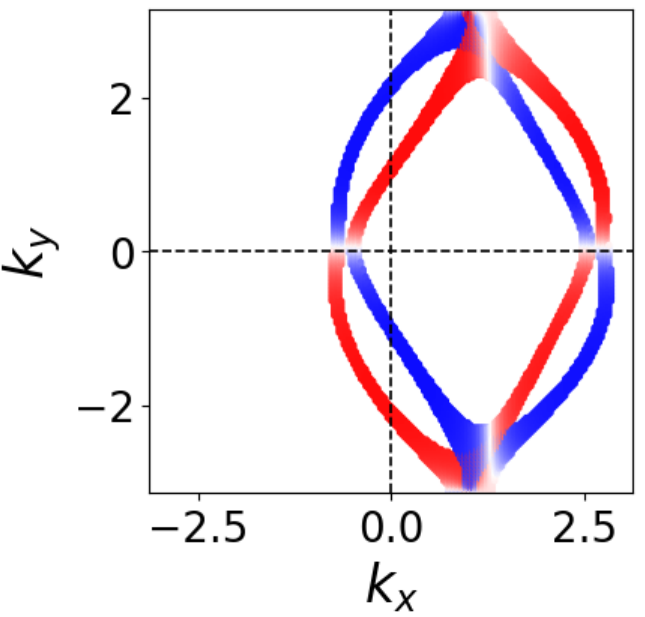}
\includegraphics[width=0.27\linewidth]{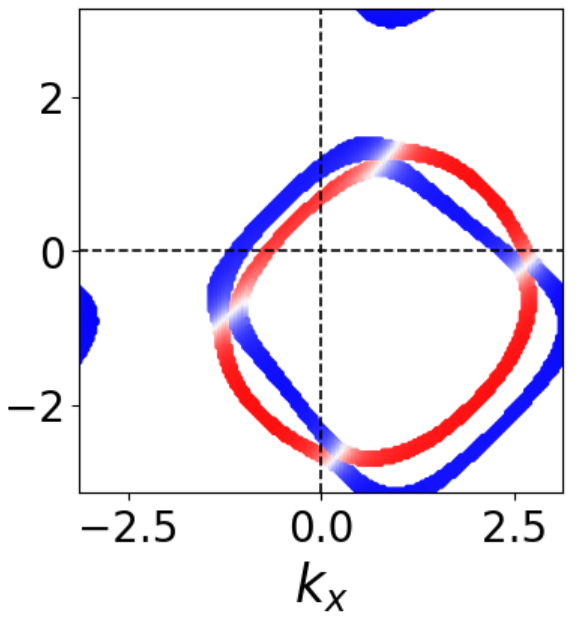}
\includegraphics[width=0.27\linewidth]{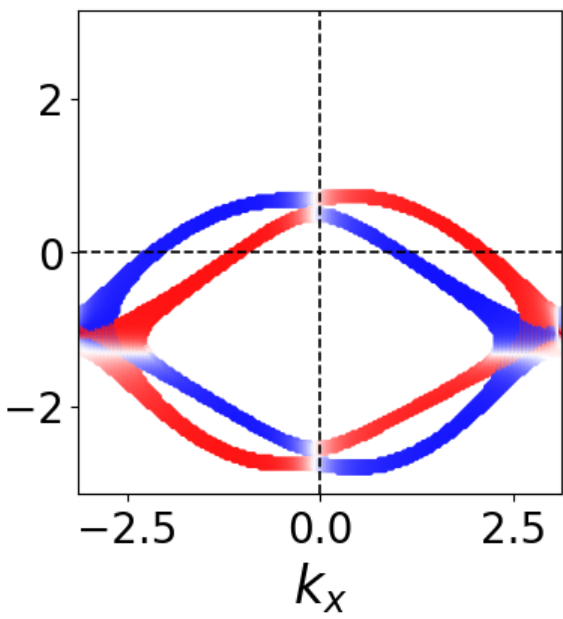}
\includegraphics[width=0.09\linewidth]{fs_plots/colorbar.pdf}   
    
    \caption{\textbf{BCL-induced spin-resolved Fermi surface.} The Fermi surfaces for \(d_{x^2-y^2}\)-wave altermagnet (top-panel) and \(d_{xy}\)-wave altermagnet (bottom-panel) for \(\alpha=0\) (left-panel), \(\alpha=\pi/4\) (middle-panel) and \(\alpha=\pi/2\) (right-panel) for filling \(f=0.30\), and other parameters set to \(t=1.0, t_j=0.30,\omega=3\) and \(\lambda=0.1\). The shape, position, and the characteristics of the Fermi surface can be directly controlled by the BCL.}
    
    \label{fig: control over Fermi surface}
\end{figure}

\textit{Light modulated Fermi surface and spin-texture.} Having seen the BCL-induced tunability of band-topology, Berry curvature and anomalous Hall response in \(d\)-wave altermagnets, now we turn to the controllability of Fermi surface and spin-textures for this system. A \(d\)-wave altermagnet without SOC is characterized by a typical spin-split Fermi surface with four-fold rotational symmetry that reflects the \(R_{\pi/2}T\)-symmetry of the altermagnetic Hamiltonian. The presence of SOC lifts the degeneracy of up and down spins at four discrete points where their respectiveFermi surfaces would intersect, see Ref.~\cite{KnolleTubaleTopology}. The same symmetry is also reflected in the spin-textures.

\begin{table}[b]
\centering
\caption{\textbf{BCL relative phase, \(\bm \alpha\), assisted modulations of nearest-neighbor (NN) and next nearest neighbor (NNN) form factors.} Here we have chosen \(\eta=2.\)}
\label{tab: modulations}
\begin{tabular}{c c c}
\hline
\hline
$\bm \alpha$ & \textbf{NN modulations} & \textbf{NNN modulations} \\
\hline
\hline
$n\pi$ & $f_0^x>f_0^y$ & $f_0^+=f_0^-$ \\
$n\pi+\pi/4$ & $f_0^x=f_0^y$ & $f_0^+<f_0^-$ \\
$n\pi+\pi/2$ & $f_0^x<f_0^y$ & $f_0^+=f_0^-$ \\
$n\pi+3\pi/4$ & $f_0^x=f_0^y$ & $f_0^+>f_0^-$ \\
\hline
\hline
\end{tabular}
\end{table}

In the presence of light irradiation, however, the spin-textures can be manipulated in a variety of ways because BCL allows control over the rotational symmetry of the system~\cite{electricPolarizationWithBCL}. We note that to understand the effect of BCL on the overall symmetries reflected in the Fermi surface as well as in the spin-textures, it suffices to analyze the \(H_0(\bm k)\) part of \(H_F(\bm k)\). This is because, in the high frequency regime, the other parts in \(H_F(\bm k)\) are suppressed as \(\omega^{-m},\text{ for } m>0\) and can be thought of as a perturbation over \(H_0(\bm k)\). It turns out that \(H_0(\bm k)\) can be obtained from the original \(H(\bm k)\) by a simple set of substitution rules for the form factors, \(\cos k_\alpha \rightarrow f_0^\alpha \cos (k_\alpha+k_{0\alpha})\), where \(\alpha\in \{x,y,+,-\}\) and \(k_\pm = k_x\pm k_y\) (\(k_{x,y}\) appear in the nearest-neighbor (NN) form factors and \(k_\pm\) appear in the next-nearest-neighbor (NNN) form factors). Similar substitutions can be made for $\sin$ terms. The modulations \(f_0^\alpha\) and shifts \(k_{0\alpha}\) are ultimately dependent upon the time-average (over a time-period) of the cosines and sines of various components of the vector potential. Directional anisotropy in modulations, \(f_0^x\neq f_0^y\) will necessarily imply breaking of four-fold rotation symmetry of the (effective) NN hopping terms and \(f_0^+\neq f_0^-\) will likewise imply breaking of the four-fold rotation symmetry of the (effective) NNN hopping terms. On the other hand, \(k_{0\alpha}\) can be thought as an \emph{effective static \(U(1)\) gauge field} or an \emph{effective twisted boundary condition} in the real space. For monochromatic CL or even elliptically polarized light (EPL), however, we always have \(k_{0\alpha}=0.\) The same is also true for BCL with an odd frequency ratio \(\eta\) between the two beams. For even \(\eta\), the NN shift vector \(\bm k_0 = (k_{0x},k_{0y})\) and the NNN shift vector \(\bm k_0' = 0.5(k_{0+}+k_{0-},k_{0+}-k_{0-})\) are generically non-vanishing and can be rotated in the \(k_x-k_y\) plane by tuning the relative phase \(\alpha\) of the BCL. Moreover, we find that the directions of rotation are opposite for \(\eta\text{ mod } 4 = 2\) and \(\eta\text{ mod } 4 = 0\).

There is no directional anisotropy of modulations for CL, while for EPL the anisotropy directly depends on the orientation of the semi-major axis of the ellipse, with a vanishing anisotropy for certain symmetric orientations. For BCL with odd \(\eta\), we find that there is an anisotropy for \(\eta\text{ mod 4}=1\), but no anisotropy for \(\eta\text{ mod 4}=3\), independent of \(\alpha\). For even \(\eta\), the anisotropies are directly controlled by \(\alpha\) and show distinctive behavior for effective NN and NNN hopping terms, as summarized in Table~\ref{tab: modulations}. Keeping in mind that the non-altermagnetic part of the Hamiltonian for both the \(d\)-wave altermagnets originates from NN hopping, while the altermagnetic part of \(d_{xy}\)-wave altermagnet originates from NNN hopping and that of \(d_{x^2-y^2}\)-wave altermagnet originates from NN hopping, the results for Fermi surface in Fig.~\ref{fig: control over Fermi surface} and spin-texture evolution~\cite{SM1} as a function of \(\alpha\) can be explained by our discussion above. We particularly highlight the cases \(\alpha=n\pi/2\) for \(d_{xy}\)-wave altermagnet, when the four-fold rotation of non-altermagnetic part is broken while \(R_{\pi/2}T\) is intact in the altermagnetic part, giving rise to a spin-split Fermi surface with broken \(R_{\pi/2}T\) but still a non-zero net magnetization. However, the \(\sigma_z\) spin-texture entirely captures the symmetry of the altermagnetic part of the effective Hamiltonian of the both models. When \(R_{\pi/2}T\) is broken in the altermagnetic part, it gives rise to a weak ferromagnetism leading to non-vanishing spin-polarization of the filled Fermi sea. This highlights a direct control of the spin-texture via the relative phase between the two beams of the BCL, beyond the monochromatic case.

\textit{Conclusions and outlook.}
In summary, we discovered that BCL provides versatile control over altermagnetic properties. We showed that BCL allows for enhanced tunability of the band topologies and topological phase transitions as compared to single frequency drives. The various topological phase transitions that we uncovered imprint signatures on the anomalous Hall conductivity that can be measured experimentally. We presented the effect of BCL on various symmetries of the model, which can also be controlled by tuning various parameters of light. These changes are directly reflected in the shape and position of the Fermi surfaces as well as on the underlying spin-textures. The controllability of the Fermi surface via BCL can have intriguing effects on the charge-to-spin conversion ratios or on finite momentum superconductivity, among other effects. Overall, our results open new avenues for Floquet engineering of altermagnets using structured light.

\textit{Acknowledgments.} A.N. acknowledges support from the DST MATRICS grant (MTR/2023/000021). M.G. thanks SERB, India, for financial support through Project No. JBR/2020/000043.

\bibliographystyle{apsrev4-1}
\bibliography{bibliography}

\end{document}